\documentclass[english]{lni}

\usepackage[utf8]{inputenc} 
\usepackage{array}
\usepackage[T1]{fontenc}    
\usepackage{booktabs}       
\usepackage{amsfonts}       
\usepackage{graphicx}
\usepackage{amsmath}
\usepackage[shortlabels]{enumitem}

\begin{document}
    
\title[Stylish RLAs]{Stylish Risk-Limiting Audits in Practice}

\author[A.K. Glazer \and J.V. Spertus \and
P.B. Stark]{%
Amanda K.\ Glazer\footnote{Authors listed alphabetically. Authors contributed equally to this work.}\footnote{%
Department of Statistics, University of California, Berkeley, CA, USA
\email{amandaglazer@berkeley.edu}
} \and
Jacob V.\ Spertus\footnote{%
Department of Statistics, University of California, Berkeley, CA, USA
\email{jakespertus@berkeley.edu}
}
\and
Philip B.\ Stark\footnote{%
Department of Statistics, University of California, Berkeley, CA, USA
\email{pbstark@berkeley.edu}
}}

\startpage{1} 
\editor{TBD} 
\booktitle{Proceedings of E-Vote-ID 2023} 

\maketitle

\begin{abstract}
   Risk-limiting audits (RLAs) can use information about which ballot cards contain which contests 
   (\emph{card-style data}, CSD) to ensure that each contest receives adequate scrutiny, without examining more cards than necessary. 
   RLAs using CSD in this way can be substantially more efficient than RLAs that sample indiscriminately from all cast cards.
   We describe an open-source Python
   implementation of RLAs using CSD for the Hart InterCivic Verity voting system and the Dominion Democracy Suite\textsuperscript{\textregistered} 
   voting system.
   The software is demonstrated using 
   all 181~contests in the 2020 general election and all 214~contests in the 2022 general election in Orange County, CA, USA, the fifth-largest
   election jurisdiction in the U.S., with over 1.8~million active voters.
   (Orange County uses the Hart Verity system.)
   To audit the 181~contests in 2020 to a risk limit of 5\% without using CSD would have required a complete hand tally of all 3,094,308 cast ballot cards.
   With CSD, the estimated sample size is about 20,100 cards, 0.65\% of the cards cast---including one tied contest that required a complete hand count.  
   To audit the 214~contests in 2022 to a risk limit of 5\% without using CSD would have required a complete hand tally of all 1,989,416 cast cards.
   With CSD, the estimated sample size is about 62,250 ballots, 3.1\% of cards cast---including three contests with margins below 0.1\% and 9 with margins below 0.5\%.
\end{abstract}


\section{Introduction}
Risk-limiting audits (RLAs) manually inspect ballots from a trustworthy record of the votes\footnote{%
  Not all paper vote records are trustworthy. See, e.g., \cite{appelEtal20,appelStark20}.
  Absent a trustworthy record of the vote, no audit can provide affirmative evidence that the reported winners really won.
} 
to provide affirmative evidence that electoral outcomes (i.e., who won, not the exact vote counts) are correct if they are indeed correct, and (with a prespecified minimum probability) to correct any outcomes that are wrong.
The maximum chance that an RLA does not correct a result that is wrong is the \textit{risk limit}.
For example, an RLA with a risk limit of 5\% guarantees that if the reported outcome is wrong, the audit has at least a 95\% chance of catching and correcting the reported outcome before it is certified.
When the outcome is correct, RLAs may inspect only a small fraction of all ballot cards, saving considerable labor compared to a full manual recount.

According to the 2018 National Academies of Science, Engineering, and Medicine report \emph{Securing the Vote: Protecting American Democracy} \cite[Recommendation 5.8]{nasem18}:
\begin{quote}
States should mandate risk-limiting audits prior to the certification of election results. 
With current technology, this requires the use of paper ballots. 
States and local jurisdictions should implement risk-limiting audits within a decade. They should begin with pilot programs and work toward full implementation. 
Risk-limiting audits should be conducted for all federal and state election contests, and for local contests where feasible.
\end{quote}
No jurisdiction currently mandates RLAs of every contest in every election, or even every federal and statewide contest.
For example, Georgia law only requires auditing one contest every two years,
and Colorado law requires auditing two contests in each election.
While some officials claim that such sparse or infrequent auditing shows that their voting systems work flawlessly,\footnote{%
See, e.g., Georgia Secretary of State Brad Raffensperger's claims that the audit of one contest in 2020 ``reaffirmed that the state's new secure paper ballot voting system accurately counted and reported results.'' \url{https://sos.ga.gov/news/historic-first-statewide-audit-paper-ballots-upholds-result-presidential-race} (last visited 2~May 2023)
and that the audit of one contest in 2022
``shows that our system works and that our county election officials conducted a secure, accurate election.'' \url{https://sos.ga.gov/news/georgias-2022-statewide-risk-limiting-audit-confirms-results} (last visited 2~May 2023)
}
auditing one reported outcome says nothing about whether any other reported outcome: every contest should get some scrutiny
(or at least have a high probability of being audited).

Historically, auditing local contests together with jurisdiction-wide contests using a single audit sample has been infeasible.
Indeed, when some contests are small and others are jurisdiction-wide, RLA methods that sample ballots uniformly at random would require a full hand count throughout the jurisdiction, even when every margin (as a percentage of votes in the contest) is large.

However, \cite{glazerEtal21} presented an approach to RLA sampling that allows many contests of different sizes to be audited efficiently using the same sample.
Instead of sampling cards uniformly at random, 
the method uses \emph{card-style data} (CSD) 
and \emph{consistent sampling} to ensure that each contest gets the scrutiny it needs, without entailing unnecessary scrutiny of other contests.
They illustrated their method with simplified examples involving only two contests, but in the U.S., there can be hundreds of contests in a single election.

We incorporated the \cite{glazerEtal21} method into the SHANGRLA Python RLA library,\footnote{%
\url{https://github.com/pbstark/SHANGRLA}
} 
leveraging recent developments in formulating RLAs as hypothesis tests about the means of bounded, finite lists of numbers \cite{stark20a} and efficiently measuring risk using test supermartingales \cite{spertus23, stark23b, waudby-smithEtal21}. 
To illustrate the practical implications of CSD, we applied the method to historical data from the 2020 and 2022 general elections in Orange County, CA, which comprised 181~contests and 214~contests, respectively.
In both elections, standard RLA methods would have required a full hand count to audit every contest to a risk limit of 5\%.
The new method reduces the estimated audit workload by more than 99\% for the 2020 election and by 97\% for the 2022 election.

The next section reviews terminology, describes the problem, and summarizes the building blocks, including simultaneous card-level comparison audits of multi-style elections. 
Section~\ref{sec:software} provides a high level description of our software.
Section~\ref{sec:oc} describes the 2020 and 2022 Orange County elections, gives an overview of our implementation, and presents sample size estimates for RLAs with and without CSD. 
Code that produced our results is available at \url{https://github.com/pbstark/SHANGRLA}.
Section~\ref{sec:discussion}, discusses ramifications for real-world RLAs and provides recommendations for practice.

\section{Background}
\label{sec:alg-choices}

In the U.S., a \emph{ballot} consists of one or more \emph{cards}, individual pieces of paper.
(U.S.\ elections often contain more contests than can be printed on a single piece of paper in a readable font.)
Each card has a \emph{style}, which for our purposes is the collection of contests on that card.
Because ballot boxes are generally designed so that the cards do not land in the order in which they were cast, it is typically impossible to reassemble a ballot from its component cards once it has been cast.
Thus cards, not ballots, are the atomic sampling unit in RLAs.

When ballots have multiple cards, no contest is on more than half the cards.
Contests that are not jurisdiction-wide are on even fewer cards.
Following the terminology of \cite{stark23b}, the \textit{sampling domain} of a contest is the population from which cards are sampled in an RLA.
For the RLA to be valid, the sampling domain for a given contest generally must include every card that contains that contest.
In practice, the sampling domain for RLAs has been either all cards cast in the election, or just the cards containing a particular contest. 
When the sampling domain for a contest includes cards that do not contain the contest, the audit generally needs to examine more ballots
(when the outcome is correct) than it  would
if the sampling domain were limited to cards containing the contest.

In particular, audits that directly check the voting system's interpretation and tabulation of votes are more efficient when the sampling domain is limited to cards that contain the contest because
the error rate (per card) required to alter the outcome is smaller the larger the denominator (the sampling domain) is.
Testing whether the error rate is below a small threshold requires more data than testing whether it is below a larger threshold. 

The \emph{diluted margin}, the margin in votes divided by the number of cards in the sampling domain for the contest, captures this phenomenon.
Smaller diluted margins lead to larger audit sample sizes; expanding the sampling domain increases the ``dilution,'' reducing the diluted margin. 

\subsection{Card-level Comparison Audits and Card-Style Data}

RLAs can use data from voting systems and from manually inspected cards in a number of ways. 
RLAs that check for error by comparing ballots to their machine interpretations are called \textit{comparison} audits; those that check outcomes without relying on the voting system's interpretations are \textit{polling} audits. 
Furthermore, RLAs may sample and check vote totals for \textit{batches} of ballot cards---typically machines or precincts---or individual cards. 
Adopting the terminology of \cite{stark23c}, we
refer to audits that sample individual cards and compare a human reading of the votes on each sampled card to the CVR for that card as \textit{card-level} audits.
The literature sometimes often
calles these \emph{ballot-level}, but card-level
is more accurate nomenclature because
CVRs are generally for individual cards, and ballots comprise more than one card in many jurisdictions.

All else equal, card-level audits are more efficient than batch-level audits; comparison audits are more efficient than polling audits; and \textit{card-level comparison audits} are the most efficient approach. In a card-level comparison audit, the estimated sample size scales with the reciprocal of the diluted margin.

To conduct a card-level comparison audit, the voting system must produce \emph{cast-vote records} (CVRs)---the system's record of the votes on each card.
(However, see \cite{stark23c},
which shows how to conduct a card-level comparison audit using
``overstatement-net-equivalent'' CVRs derived from batch-level results.)
Moreover, there must be a known 1:1 mapping from the physical card to its particular CVR; some voting systems cannot provide such a link, or cannot provide a link without compromising voter privacy.
That link might be provided by exporting CVRs in the order in which the cards were scanned and keeping the cards in that order, or by imprinting a number on each card before or as it is scanned, and including the imprinted number in the CVR for that ballot or as part of the filename of the CVR.
\cite{glazerEtal21} showed that an audit can rely on CVRs to infer CSD: consider a card to contain a contest if the CVR for that card contains the contest.
Even though the CVRs might be inaccurate or incomplete (otherwise, no audit would be needed), their method ensures that errors in CSD derived from CVRs do not compromise the risk limit.
CSD makes it possible to minimize the sampling domain (maximizing the diluted margin) for each contest, considerably lowering audit workloads when
contest outcomes are correct. 

\cite{glazerEtal21} also showed how to combine CSD with \emph{consistent sampling} \cite{rivest2018}, which ensures that cards drawn for the purpose of auditing a given contest can also be used in the audit of other contests that appear on the sampled cards.
Exploiting such overlap further reduces the estimated workload.
If the voting system does not provide CVRs linked to cards, CSD can be derived by manually sorting the cards (a very labor intensive alternative), or by processing them in homogeneous batches in the first place.
That is straightforward for precinct-based voting systems,
but many jurisdictions do not sort cards by style before scanning them.

When CSD are derived from CVRs, the RLA can also use those CVRs for card-level comparison auditing, which is much more efficient than ballot-polling or batch-level comparison auditing.
For this reason, the audits we describe in the remainder of this paper are card-level comparison audits, but the software also supports ballot-polling audits with CSD.
We now describe how the software is implemented to run a card-level comparison audit.

In broad brush, the procedure imports audit parameters (such as risk limits, risk-measuring functions, and the strategy for estimating the initial sample size), election data (including the reported winners, the CVRs, and upper bounds on the number of cards cast in each contest\footnote{%
All RLAs require information that a good canvass should generate routinely, including upper bounds on the number of validly cast cards 
that contain each contest,
which can be derived from registration data, voter participation data, and physical inventories of ballot cards.
Absent that information, even a ``full'' hand count is meaningless, since there is no way to know whether the count includes every validly cast ballot.}), and contest-specific data (such as candidate names, and the social choice function).

CSD is inferred from the CVRs.
The CVRs are checked for consistency with the other inputs.
An initial sample size is determined for each contest, which implies a sampling probability
for each card that contains the
contest.
The probability that a given card is sampled is the largest sampling probability for each audited contest on the card.
Summing those maximum probabilities across cards
is an estimate of the total initial sample size.
A sample is drawn using \emph{consistent sampling};
the corresponding cards are retrieved and interpreted manually; the resulting \emph{manual vote records} (MVRs) are imported; the attained risk for each audited contest is calculated; and 
any contests for which the risk limit has been attained or for which there has been a full hand count are 
removed from future auditing rounds.
If every audited contest has been removed, the audit stops;
otherwise, a next-round sample size is determined for the 
remaining contests, and the process repeats.

In more detail, the algorithm is 
as follows (adapted from \cite{glazerEtal21}):
\begin{enumerate}[topsep=0pt, partopsep=0pt, itemsep=2pt,parsep=2pt]
    \item Set up the audit
    \begin{enumerate}[topsep=0pt, partopsep=0pt, itemsep=2pt,parsep=2pt]
        \item Read contest descriptors, 
        candidate names,
        social choice functions,
        upper bounds on the number of cards that contain each contest,
        and reported winners.
        Let $N_c$ denote the upper bound on the number of cards that contain contest $c$, $c=1, \ldots, C$.
        \item Read
        audit parameters (risk limit
        for each contest, risk-measuring function to use for each contest, assumptions about errors for computing initial sample sizes), and seed for pseudo-random sampling.
        \item Read ballot manifest.
        \item Read CVRs.
    \end{enumerate}
    \item Pre-processing and consistency checks
    \begin{enumerate}[topsep=0pt, partopsep=0pt, itemsep=2pt,parsep=2pt]
    \item Check that the winners according to the CVRs are the reported winners.
    \item If there are more CVRs that contain any contest than the upper bound on the number of cards that contain the contest, stop: something is seriously wrong.
    \item If the upper bound on the number of cards that contain any contest is greater than the number of CVRs
    that contain the contest, create a corresponding set of ``phantom'' CVRs as described in section~3.4 of
    \cite{stark20a}.
    The phantom CVRs are generated separately for each contest: each phantom card contains only one contest.
    \item If the upper bound $N_c$ on the number of cards that contain contest $c$ is greater than the number of physical cards whose locations are known, create enough ``phantom'' cards to make up the difference.
    \end{enumerate}
    \item Prepare for sampling
    \begin{enumerate}[topsep=0pt, partopsep=0pt, itemsep=2pt,parsep=2pt]
    \item Generate a set of SHANGRLA \cite{stark20a} assertions $\mathcal{A}_c$ 
    for every contest $c$ under audit. 
    \item Initialize $\mathcal{A} \leftarrow \cup_{c=1}^C \mathcal{A}_c$
    and $\mathcal{C} \leftarrow \{1, \ldots, C\}$.
    \item Assign independent uniform pseudo-random numbers to
    CVRs that contain one or more contests under audit (including ``phantom'' CVRs), using a high-quality PRNG \cite{OttoboniStark19}.
    (The code uses cryptographic quality pseudo-random integers uniformly distributed on $0, \ldots, 2^{256}-1$.)
    Let $u_i$ denote the number assigned to CVR $i$.
  \end{enumerate}
  \item Main audit loop.
  While $\mathcal{A}$ is not empty: 
    \begin{enumerate}[topsep=0pt, partopsep=0pt, itemsep=2pt,parsep=2pt]
        \item Pick the (cumulative) sample sizes 
        $\{S_c\}$ for $c \in \mathcal{C}$ to attain by the end of this round of sampling.
        The software offers several options for picking $\{S_c\}$, including some based on simulation.
        The desired sampling fraction $f_c := S_c/N_c$ for contest $c$ is the sampling probability for each card that contains contest $k$, treating cards already in the sample as having sampling probability 1.
        The probability $p_i$ that previously unsampled card $i$ is sampled in the next round is the largest of those probabilities:
        $p_i := \max_{c \in \mathcal{C} \cap \mathcal{C}_i} f_c $,
        where $\mathcal{C}_i$ denotes the contests on card $i$.
        \item Estimate
        the total sample size to be $\sum_i p_i$, where the sum is across all cards except phantom cards.
        \item Choose thresholds $\{t_c\}_{c \in \mathcal{C}}$ so that $S_c$ ballot cards containing contest $c$ have a sample number $u_i$ less than or equal to $t_c$.
        \item Retrieve any of the corresponding ballot cards that have not yet been audited and inspect them manually to generate MVRs.
        \item Import the MVRs.
        \item For each MVR $i$:
        \\
           ${}$~
           For each $c \in \mathcal{C}$:\\
            ${}$ \hspace{0.2cm}
            If $u_i \le t_c$, then for each $a \in \mathcal{A}_c \cap \mathcal{A}$:
            \begin{itemize}
            \item If the $i$th CVR is a phantom,
               define $a(\mbox{CVR}_i) := 1/2$.
            \item
               If card $i$ cannot be found or if it is a phantom, define $a(\mbox{MVR}_i) := 0$.
            \item Find the overstatement of assertion $a$ for CVR $i$, 
               $a(\mbox{CVR}_i) - a(\mbox{MVR}_i)$.
            \end{itemize}
        \item  Use the overstatement data from the previous step to update the 
        measured risk for every assertion $a \in \mathcal{A}$.

        \item Optionally, conduct a full hand count of one or more contests, for instance, if the audit data suggest the outcome is wrong or if the auditors think a hand count will be less work than continuing to sample.
        
        \item Remove from $\mathcal{A}$ all assertions $a$ that have met their risk limits
        or that are for contests for which there has been a full hand count. (The audits of those assertions are complete.)

        \item Remove from $\mathcal{C}$ all contests $c$ for which $\mathcal{A}_c \cap \mathcal{A} = \emptyset$ (the audits of those contests are complete).
    \end{enumerate}
    \item Replace the reported outcomes of any contests that were fully hand counted by the outcomes according to those hand counts.
\end{enumerate}

\section{Software}
\label{sec:software}
The software can read Dominion Democracy Suite\textsuperscript{\textregistered} and Hart InterCivic Verity CVRs and manifest files.
Because file sizes in large jurisdictions can be unwieldy, the software can read compressed CVR files (.zip format containing XML records).

\begin{figure}
    \centering
    \includegraphics[width = \textwidth]{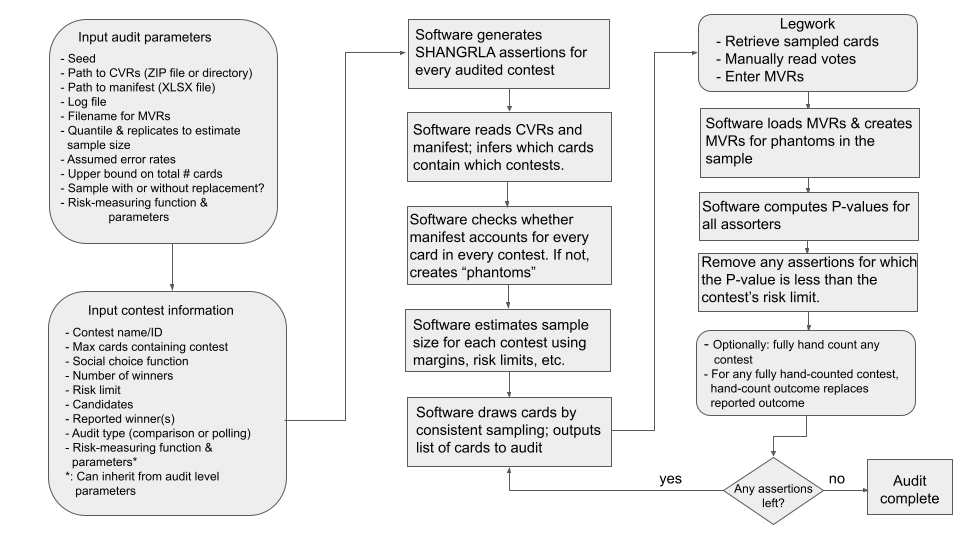}
    \caption{Workflow for simultaneous card-level comparison audit using SHANGRLA software with CSD and consistent sampling.
    Boxes with rounded corners involve inputs from the auditors.}
    \label{fig:flowchart}
\end{figure}

Figure~\ref{fig:flowchart} sketches the workflow to audit a collection of contests using CSD derived from CVRs.
The user specifies parameters of the audit and the contests to be audited, including paths to data and output files, a trustworthy upper bound on the number of cards cast (e.g., a bound from participation records, ballot accounting, pollbook reconcilation, etc.---not the voting system's own reported number of cards), contest information, risk limits, risk measuring functions and their tuning parameters (defaults are available), information used to estimate initial sample sizes (defaults are available), and whether to use CSD.

The software then constructs SHANGRLA assertions (or reads RAIRE assertions in json for IRV contests), reads CVRs and manifests, constructs ``phantom'' CVRs to account for missing cards if necessary, sets margins for overstatement assorters, estimates initial sample sizes, draws random ballots by consistent sampling, and returns their locations to the auditors. 

The auditors retrieve the indicated  cards, manually read the votes from those cards, and input the MVRs, which the audit software subsequently reads from a file.
The software uses the specified risk-measuring function(s), the CVRs, and the MVRs to compute a $P$-value for every assertion.
All assertions with $P$-values below the risk limit for their corresponding contests are considered ``confirmed.'' 
If any assertions remain unconfirmed, the software will estimate the number of additional cards to examine to confirm those assertions, draw a sample of that size, and export the identifiers of the ballot cards for the auditors to retrieve and interpret. 

This process repeats until every assertion has been confirmed or there has been a full hand count of the contest.
At any point, the auditors can choose to stop sampling at random and simply tabulate the rest of the votes
in one or more contests (e.g., if they judge that that would be more efficient, or if the audit sample indicates that the reported outcome is in fact wrong).

\section{Orange County Election Audits}
\label{sec:oc}

Orange County, CA, is the third most populous county in California (3.19~million as of the 2020 census, with over 1.81~million active voters\footnote{%
\url{https://ocvote.gov/datacentral/}, last visited 8~June 2022}).
It has more registered voters than 24~U.S.\ states, and is the country's fifth-largest election jurisdiction, after Los Angeles, CA; Maricopa, AZ; Harris, TX;  and
San Diego, CA.\footnote{%
2022 Election Administration and Voting Survey (EAVS), U.S. Election Assistance Commission, \url{https://www.eac.gov/sites/default/files/2023-06/2022_EAVS_for_Public_Release_V1.xlsx}, released 29~June 2023.
Last visited 9~July 2023.}
As of this writing, Orange County has 2204~precincts and approximately 181~voting centers. 
Orange County uses the Hart Intercivic Verity system.
The county first piloted an RLA in 2011,\footnote{%
See California Secretary of State Report to the US Election Assistance Commission, \url{https://admin.cdn.sos.ca.gov/reports/2011/post-election-audit-report-20111130.pdf}, last visited 15~May 2023.
}
conducted two additional pilots in 2018\footnote{%
See \url{https://verifiedvoting.org/wp-content/uploads/2020/08/2018-RLA-Report-Orange-County-CA.pdf}, last visited 15~May 2023.
}
and seven pilots between 2020 and 2022 mandated by
California Elec. Code, \S\S~15365--15367.

In this paper, we use data from the November 2020 and 2022 General Elections.
We estimate the number of cards that would need to be inspected for an RLA with 5\% risk limit, with and without style information.
(While many states that require or authorize RLAs do not specify a risk limit in statute, 5\% is a common value in practice. 
It was the statutory requirement in California's pilot program, Cal. Elec. Code \S~15367.
Sample sizes for RLAs generally scale approximately like the log of the risk limit, so sample sizes for a risk limit of 1\% would be about $\log(0.01)/\log(0.05)-1 \approx 54\%$ larger.)
Table~\ref{tab:oc-summary} summarizes these elections and the results of our calculations
for all contests, and for all contests with margins greater than $0.1\%$, $0.5\%$, or $1\%$.
Section~\ref{sec:recounts}
that because of automatic manual recount laws
in various states, it may make
sense to omit contests with small margins from the workload estimates.

To audit cross-jurisdictional contests requires sampling from all cards cast in the contest, not just those cast in one jurisdiction.
Since we did not have access to CVRs for other counties, the sample size estimates we report
treat every contest in both elections as if it were entirely contained in Orange County.
In particular, the estimates take the margins of statewide contests to be the margins within Orange County alone,
and ignore the fact that the resulting audit burden would be shared across all jurisdictions with voters eligible to vote in those contests.
The results still give a reasonable estimate of the workload required to audit a large number of (partially) overlapping contests simultaneously, and it is generally the \emph{smaller} contests that drive audit workload for audits that do not use CSD, for reasons explained above.
In particular, statewide contests appear on every ballot in each jurisdiction and on approximately the same fraction of cards in each jurisdiction (depending on the number of local contests in each jurisdiction).
Moreover, because Orange County has more registered voters than 24~U.S.\ states, it is a reasonable proxy for many statewide audits.

The actual sample size depends on the luck of the draw---which particular cards end up in the sample---and on the errors in the CVRs for those cards.
We estimate sample sizes using two assumed error rates: no error at all, and one 1-vote overstatement per 1,000 cards, i.e., a rate of $10^{-3}$.
(A one-vote overstatement occurs if the CVR has an error that increased the margin of a reported winner over a reported loser by one vote, e.g., if the card shows a valid vote for a loser but the CVR shows an undervote or overvote, or if the card shows an overvote, but the CVR shows a valid vote for a reported winner.)
When CVRs are error-free, the sample size is deterministic.
For the assumed rate of $10^{-3}$,
we generated artificial data that reflects a one-vote overstatement error every 1,000 ballots, starting with an error in the first CVR in the sample. 
The risk-measuring function is the ALPHA supermartingale \cite{stark23b}, with the \texttt{optimal\_comparison} estimator
of \cite{spertus23}.
That estimator depends on an assumed rate of 2-vote overstatement errors in the CVRs, which we set to $10^{-4}$.

\begin{table}[ht!]
    \centering
    \begin{tabular}{r|l|r|r|r|r}
     & \textbf{Year} & \multicolumn{2}{c|}{\bf{2020}} & \multicolumn{2}{c}{\bf{2022}} \\
    \hline
    1 & \textbf{Turnout} & \multicolumn{2}{c|}{1,546,570} & \multicolumn{2}{c}{994,227} \\
    2 & \textbf{Cards cast} & \multicolumn{2}{c|}{3,094,308} & \multicolumn{2}{c}{1,989,416} \\
    3 & \textbf{Total contests} &  \multicolumn{2}{c|}{181} & \multicolumn{2}{c}{214} \\
    \hline
    4 & \textbf{Exact ties} & \multicolumn{2}{c|}{1} & \multicolumn{2}{c}{0} \\
    5 & \textbf{Margins below 0.1\%} & 
    \multicolumn{2}{c|}{1} & \multicolumn{2}{c}{3} \\
    6 & \textbf{Margins below 0.5\%} & \multicolumn{2}{c|}{4} & \multicolumn{2}{c}{9} \\
    7 & \textbf{Margins below 1.0\%}& \multicolumn{2}{c|}{5} & \multicolumn{2}{c}{14} \\
    \hline 
     & \textbf{Sample sizes} & \multicolumn{4}{c}{\textbf{rate of 1-vote overstatements}} \\
    \cline{3-6}
    & & 0 & $10^{-3}$ & 0 & $10^{-3}$ \\
    \cline{2-6}
    8 & \bf{all contests} & 20,112 & 37,996 & 62,251 & 119,814 \\
    9 & & (0.6\%) & (1.2\%)  & (3.1\%) & (6.0\%) \\ 
    \hline
    10 & \bf{omit margins} $\leq$0.1\% & 15,964  & 33,852 & 22,110 & 33,215  \\
    11 & & (0.5\%) & (1.1\%) & (1.1\%) & (1.7\%)\\ 
    \hline
    12 & \textbf{omit margins $\leq$0.5\%} &  9,228 & 11,347 & 11,053 & 14,125  \\
    13 & & (0.3\%) & (0.4\%) &  (0.6\%) & (0.7\%) \\ 
    \hline
    14 & \textbf{omit margins $\leq$1\%} & 7,827 & 9,634 & 8,123  & 9,980   \\
    15 &  & (0.3\%) & (0.3\%) & (0.4\%)  &  (0.5\%) \\ 
    \end{tabular}
    \vspace{5mm}
    \caption{Summary of the 2020 and 2022 General Elections in Orange County, CA.
    Row~4 is the number of contests reported to be tied.
    Rows~5--7 are the number of contests with reported margins
    below 0.1\%, 0.5\%, and 1\%, respectively.
    Rows 8--15 are the sample sizes to confirm all contests to a risk limit of 5\%, expressed as the number of cards (rows 8, 10, 12, 14) or the percentage of all  cards (in parentheses, rows 9, 11, 13, 15),
    when the audit finds no errors, or when the rate of one-vote overstatement errors is 1 in 1,000 CVRs.
    (A one-vote overstatement occurs if correcting the error reduces the margin between a reported winner and a reported loser by one vote, e.g., if the CVR erroneously counts a vote for the reported loser as an undervote.)
    When there is no error, the sample size is deterministic.
    When there are errors, the sample size depends not only on their rate, but on the order in which they occur.
    To simplify the calculations,
    we estimate the sample size by assuming that the first CVR shows an error, and thereafter errors are equispaced, one every 1,000 ballots.
    Rows~8 and 9 are for all contests, including tied contests.
    Rows~10 and 11 exclude contests with reported margins less than or equal to 0.1\%, a threshold some states use for automatic recounts (see section~\ref{sec:recounts}).
    Rows~12 and 13 exclude contests with margins less than or equal to 0.5\%, a common threshold for automatic recounts.
    Rows~14 and 15 exclude contests with margins less than or equal to 1\%, another common automatic recount threshold.    For the purpose of illustration, the calculations assume that every contest (including statewide contests) is entirely contained in Orange County.}
    \label{tab:oc-summary}
\end{table}

\subsection{Audits and Recounts} \label{sec:recounts}

If a jurisdiction conducts a  \emph{manual} recount of the ballots after a robust canvass, there is no need for an RLA (some states allow machine recounts).
Many U.S.\ states conduct automatic recounts for contests with small reported margins.
Alabama, Arizona, Colorado, Connecticut, Delaware, Florida, New York, Ohio, Pennsylvania, and Washington recount contests with margins less than $0.5\%$
(possibly with exceptions).\footnote{%
\url{https://ballotpedia.org/Election_recount_laws_and_procedures_in_the_50_states}, last visited 2~July 2023.
Washington's automatic recounts are machine recounts, not hand recounts, so they do not obviate the need for an RLA.} 
Hawaii automatically recount if the margin is below $0.25\%$.
Nebraska and Wyoming have automatic recounts if the margin is less than 1\% of the winner's tally.
New Mexico and North Dakota automatically recount elections with margins less than 1\%, 0.5\%, or 0.25\%, depending on the office.
Ohio has thresholds of 0.5\% and 0.25\%, depending on the office.
Oregon has a threshold of 0.2\%.
South Carolina has a 1\% threshold.
Alaska, Montana, South Dakota, Texas, and Vermont automatically recount tied elections.
Some states have a recount threshold based on the number of votes rather than the percentage margin; for instance, Michigan has automatic recounts for statewide contests with margins below 2,000 votes.

To understand how automatic recounts affect audit workloads, we estimate the number of ballots to inspect to audit all contests regardless of their margins, and all contests with reported margins greater than 0.1\%, 0.5\%, and 1\%.

\subsection{November 2020}
In the November 2020 general election in Orange County, a total of 1,546,570 ballots (and 3,094,308 ballot cards) were cast in 181 contests.
One contest was reported to be a tie, a margin of 0 votes: in the contest for Brea Olinda Unified School District Governing Board Member, Trustee Area~5,
both Lauren Barnes and Gail Lyons were reported to receive $1,805$ votes. 
Because the reported 
result was a tie, 
auditing this contest requires a full hand count. 
If there were no way to identify which cards contain this contest without manually inspecting the cards, a full hand count of that single contest would entail manually inspecting all 3,094,308 cards
cast in the election.
In all, 27~contests have diluted margins so small
(with respect to all cards cast) that auditing each of them would require examining more than 99\% of the cast cards, unless CSD are used.

But with CSD, auditing a contest never requires inspecting more cards than contain that contest. 
This reduces the workload substantially: the estimated workload to audit all 181~contests to a risk limit of 5\% is only 20,112 cards in all, a reduction of more than 99\%.
Without the contest with margin less than $0.1\%$, the estimated sample size drops to 15,964 cards.
Without the four contests with margins less than $0.5\%$, 
the estimated sample size drops to 9,228 cards.
Without the five contests with margins less than $1\%$, 
the estimated sample size further drops to 7,827 cards.
Table~\ref{tab:small_margin_contests_2020} lists the contests with margins under $1\%$, along with their margins and estimated sample size for that contest for a 5\% risk limit RLA using CSD.
\begin{table}
    \begin{tabular}{l|r|l|r}
         \textbf{Contest} &
         \textbf{Cards}
         & \textbf{Diluted} & \textbf{Sample}   \\ 
         & \textbf{Cast} & \textbf{Margin}&\textbf{Size} \\
         \hline
         Brea Olinda Unified School District Governing Board & 4,164 & 0 & 4,164 \\
         \hspace{0.2in} Member, Trustee Area 5 & & & \\
         City of Irvine, City Council & 129,948 & 0.2\% & 3,930\\
         City of Lake Forest Member, City Council, District 1 & 10,042 & 0.2\% & 2,843\\
        Proposition 17 & 1,546,210 &0.4\% & 1,488 \\
        City of Laguna Beach Member, City Council & 16,661 & 0.8\% & 746 \\
        South Coast Water District Director & 22,046 &0.9\% & 696 \\
        Member of the State Assembly 74th District & 277,516 & 0.9\% & 652 \\[.05in]
     \end{tabular}
    \caption{Contests with margins under 1\% in the General Election in Orange County, CA, November 2020, number of cards cast, reported diluted margin, and estimated sample size to audit each of them to a risk limit of 5\%, on the assumption that the CVRs have no errors.  
    } \label{tab:small_margin_contests_2020}
\end{table}

\begin{figure}[!ht]
    \centering
    \includegraphics[scale = 0.2]{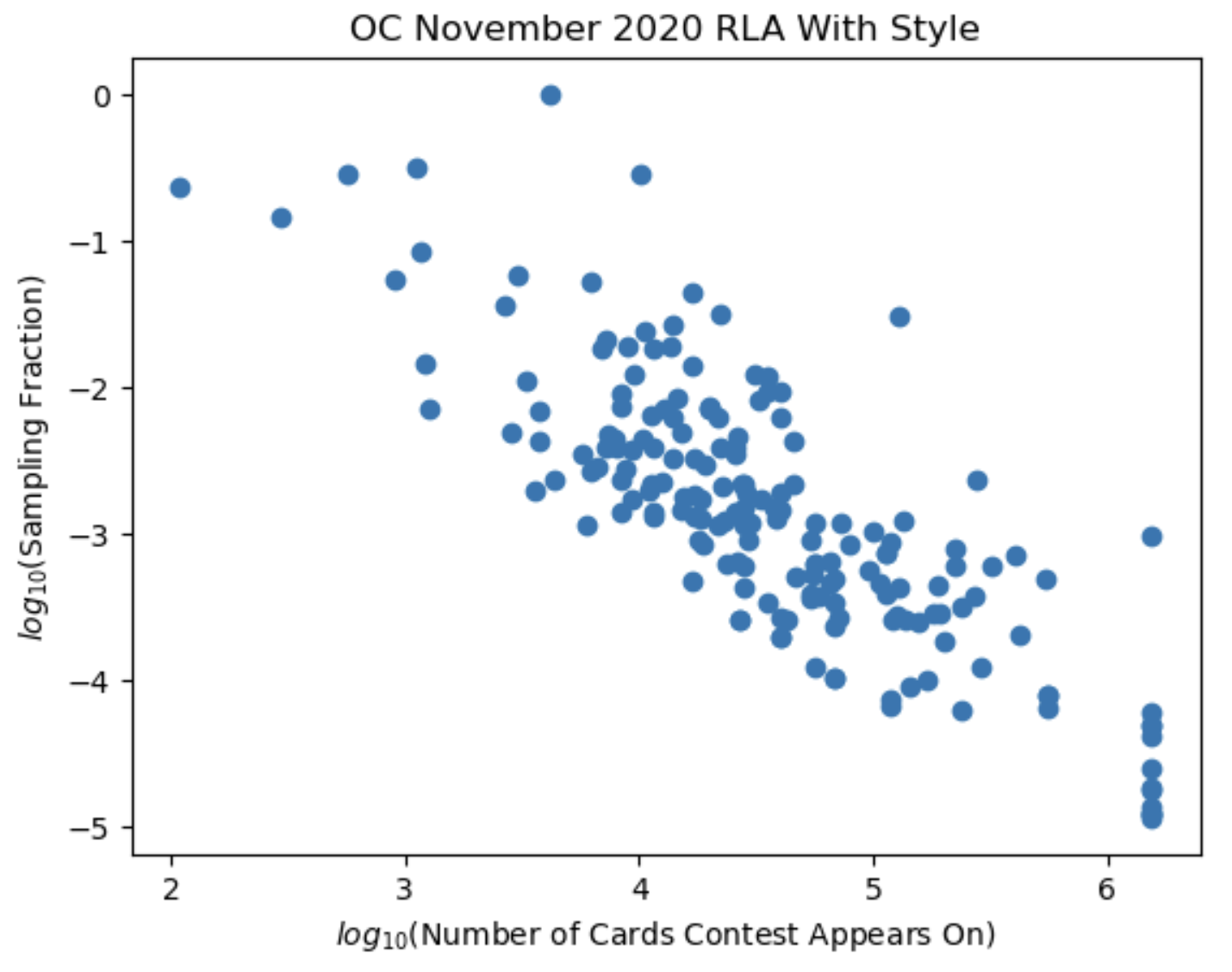}
    \includegraphics[scale = 0.2]{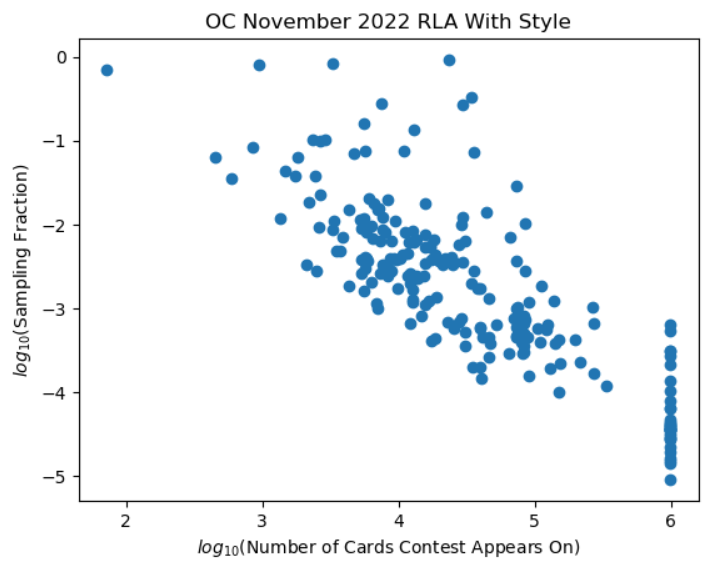}
    \caption{Log of the sampling fraction
    (cards in the sample that contain the contest, divided by cards that contain the contest) versus the log number of cards the contest appears on, for a 5\% risk limit RLA using CSD information, for
    General Elections in Orange County, CA, USA, in November 2020 (left panel, 181 contests) and 2022 (right panel, 214 contests).
    In general, for a given margin, larger contests with correct outcomes can be confirmed by examining a smaller fraction of the cards that contain the contest.
    The vertical set of points at the right edge of the plots are county-wide and statewide contests, which appear on the maximum possible number of cards.
    In 2020, all but one had a sampling fraction less than 1 in 10,000; the smallest was less than 1 in 100,000.
    In 2022, sampling fractions for the largest contests ranged from 1 in 100,000 to about 1 in 1,000. 
    \label{fig:oc2020-with}}
\end{figure}

Figure~\ref{fig:oc2020-with} shows the proportion of ballot cards containing each contest that we would expect the RLA to inspect, versus the number of cards the contest appears on (both on log scale).
In general, the sampling fraction decreases as the number of cards the contest appears on increases.

\subsection{November 2022}
In the November 2022 general election in Orange County, 994,227 ballots (comprising 1,989,416 cards) were cast in 214~contests.
Several contests had small margins.
For instance, in the vote-for-three Fountain Valley School District, Governing Board Member contest,
the margin between the winner with the fewest votes, Phu Nguyen, and the loser with the most votes, Megan Irvine, was 0.02\%.
The City of Villa Park, City Council Member contest was also multiwinner plurality with three winners; the margin between the winner with the fewest votes (Jordan Wu) and the loser with the most votes (Donna Buxton) was 0.09\%.
The margin for Measure~K in Costa Mesa, a simple majority contest, was 0.06\%.

The estimated sample size to audit all 214~contests to a risk limit of 5\% without using CSD is 1,989,415 ballot cards---essentially every card.
Indeed, there are 33~contests which, if each had been the \emph{only} contest audited, would have required inspecting more than 99\% of all cast ballot cards if CSD were not used to target the sample.

Using CSD reduces the estimated workload by 97\%: the estimated sample size to audit all 214~contests to a risk limit of 5\% is 62,251 ballot cards, about 3.1\% of the cards cast.
As mentioned above, state laws for automatic recounts typically have threshold margins of $1\%$, $0.5$\%, $0.25\%$, or $0.1$\%.
Table~\ref{tab:small_margin_contests_2022} lists the contests with margins of $1\%$ or less, their sizes, margins, and estimated sample size for a CSD RLA of each, at 5\% risk limit, computed on the assumption that the CVRs are accurate.
The right panel of Figure~\ref{fig:oc2020-with}
plots sample sizes versus contest sizes for the 214 contests.

\begin{table}[h!]
    \centering
    \begin{tabular}{l|r|l|r}
         \textbf{Contest} &
         \textbf{Cards}
         & \textbf{Diluted} & \textbf{Sample}   \\ 
         & \textbf{Cast} & \textbf{Margin}&\textbf{Size} \\
         \hline
         Fountain Valley Sch Dist Governing Board Member & 23,512 & 0.03\% & 21,772 \\
         K-City of Costa Mesa & 34,626 & 0.06\% & 11,354 \\
         City of San Clemente Member, City Council & 29,670 & 0.08\% & 7,999 \\
         City of Villa Park Member, City Council & 3,260 & 0.1\% & 2,715 \\
        Ocean View Sch Dist Governing Board Member & 35,990 & 0.2\% & 2,634 \\
        Orange Unif Sch Dist Governing Board Member, & 73,665 &  0.3\% & 2,088\\
        ${}$~ Trustee Area 4 & & & \\
         City of Westminster Member, City Council, District 1 & 7,467 & 0.3\% & 2,064\\
         La Habra City Sch Dist Governing Board Member & 12,915 & 0.3\% & 1,738 \\
         City of Los Alamitos Member, City Council, District 5 & 946 & 0.4\% & 750\\
         Rossmoor Community Services District Director & 5,540 & 0.6\% & 897 \\
         Member of the State Assembly 71st District & 85,911 & 0.7\% & 873 \\
         City of Anaheim Member, City Council, District 2 &  10,997 & 0.7\% & 835 \\
         United States Senator Full Term & 994,227 & 0.97\% & 626 \\
         City of Orange Mayor & 43,813 & 0.99\% & 612 \\
    \end{tabular}
    \vspace{0.1in}
    \caption{Contests with margins under 1\% in the General Election in Orange County, CA, November 2022, number of cards cast, reported diluted margin, and estimated sample size to audit each of them to a risk limit of 5\%, on the assumption that the CVRs have no errors. Unif Sch Dist = Unified School District.
    The sample size does not decrease monotonically as the margin grows because the sample is drawn without replacement:
    the sampling fraction matters, too.
    }
    \label{tab:small_margin_contests_2022}
\end{table}
Without the three contests with margins below $0.1\%$, the estimated sample size would be 22,110 cards.
Without the nine contests with margins below $0.5\%$, the estimated sample size would be 11,053 cards.
Without the 14 contests with margins less than 1\%, the estimated sample size would be 8,123 cards.

\section{Discussion}
\label{sec:discussion}

It is prudent to give every contest outcome some audit scrutiny: auditing some contests has little bearing on whether the outcomes of other contests are correct. 
But conducting an RLA of a large number of partially overlapping contests with a wide range of sizes has been thought to be logistically infeasible. 
By using CSD, the method of \cite{glazerEtal21} makes it practical to audit every contest in an election, which we
illustrate using data from the 2020 and 2022 general elections in Orange County, California, 
the fifth largest election jurisdiction in the U.S., with more voters than 24~entire U.S.\ states. 
(With previous methods, auditing every contest in an election is generally more challenging in larger jurisdictions than in smaller ones, because larger jurisdictions have more contests and because the small contests are on a smaller fraction of the cards cast in the jurisdiction.)

CSD sampling would reduce the workload of a 5\% risk limit RLA by more than 99\% for the 2020 election and by 97\% for the 2020 election
compared to previous approaches.
These estimates treat every contest in both elections as if they are entirely contained in Orange County.
While this is not true for statewide contests, the estimates still give an idea of the workload to audit many overlapping contests simultaneously.
These sample size estimates also assume that a card-level comparison audit
could be conducted using all validly cast cards in Orange County.
In reality, card-level comparison audits of cards cast in vote centers and polling places might require additional work, e.g., re-scanning cards centrally to create CVRs that are uniquely associated with individual cards.
Non(c)esuch \cite{stark23a} could be used to avoid such re-scanning, but would require changes to the voting equipment
to imprint nonces on cards as they are scanned.
CSD-based sampling can also be used with ONEAudit \cite{stark23c} without re-scanning or changing the voting system, albeit with some increase in sample size.
Future work will investigate the magnitude of that increase.

California law requires auditing the votes in approximately 1\% of precincts. In 2020,
Orange County's statutory audit tabulated votes on 
53,163 cards\footnote{%
\url{https://elections.cdn.sos.ca.gov/manual-tally/2020-general/orange.pdf} last visited 8~September 2023.}
and in 2022, it tabulated votes on
51,346 cards.\footnote{%
\url{https://elections.cdn.sos.ca.gov/manual-tally/2022-general/orange.pdf},
last visited 8~September 2023.
}
While it is easier to count the votes on all the cards in a precinct than to count the votes on the same number of cards selected at random,
(i)~the statutory 1\% audit does not provide evidence that outcomes are correct, 
(ii)~a CSD 5\% risk-limit RLA would have involved examining fewer ballots in all in 2020, (iii)~the CSD RLA generally involves transcribing data from fewer contests per audited card,
and (iv)~hand-counting teams generally comprise four people to tabulate votes on a single card, while comparison-audit teams generally comprise only two people.

CSD makes the recommendation of the 2018 National Academies report \cite{nasem18} practical: jurisdictions can efficiently audit every federal and state election contest as well as all local contests using samples that will generally comprise only a modest fraction of cards cast when reported outcomes are correct.
An open-source Python  implementation of the method is available at \url{https://github.com/pbstark/SHANGRLA/tree/main/shangrla}.

\subsubsection*{Acknowledgements.}
We are grateful to the Orange County Registrar of Voters and its staff, including
Justin Berardino, Roxana Castro, and Imelda Carrillo, for data, discussions, and comments on an earlier draft; and to Paul Burke for comments on an earlier draft.
This work was supported by NSF Grant SaTC–2228884.

\bibliography{pbsBib}

\end{document}